\documentclass{elsart3}
\usepackage{graphicx}
\usepackage{amssymb}
\begin{document}

\begin{frontmatter}
\title{Effect of annealing on glassy dynamics and non-Fermi liquid behavior in UCu$_4$Pd}
\author[UCR]{D.~E. MacLaughlin\corauthref{MacL}}
\corauth[MacL]{Tel. +1-951-827-5344, Fax +1-951-827-4529, e-mail: macl@physics.ucr.edu.}
\author[UCR]{M.~S. Rose}
\author[UCR]{J.~E. Anderson}
\author[CSULA]{O.~O. Bernal}
\author[LANL,JAERI]{R.~H. Heffner}
\author[KOL]{G.~J. Nieuwenhuys}
\author[UCSD]{R.~E. Baumbach}
\author[UCSD]{N.~P. Butch}
\author[UCSD]{M.~B. Maple}
\address[UCR]{Department of Physics, University of California, Riverside, CA 92521, U.S.A.}
\address[CSULA]{Department of Physics and Astronomy, California State University, Los Angeles, CA 90032, U.S.A.}
\address[LANL]{Los Alamos National Laboratory, K764, Los Alamos, NM 87545, U.S.A.}
\address[JAERI]{Japan Atomic Energy Research Institute, Tokai, Ibaraki-ken, 319-1195, Japan}
\address[KOL]{Kamerlingh Onnes Laboratory, Leiden University, 2300 RA Leiden, The Netherlands}
\address[UCSD]{Department of Physics and Institute for Pure and Applied Physical Sciences, University of California, San Diego, La Jolla, California 92093, U.S.A.}

\begin{abstract}
Longitudinal-field muon spin relaxation (LF-$\mu$SR) experiments have been performed in unannealed and annealed samples of the heavy-fermion compound UCu$_4$Pd to study the effect of disorder on non-Fermi liquid behavior in this material. The muon spin relaxation functions~$G(t,H)$ obey the time-field scaling relation~$G(t,H) = G(t/H^\gamma)$ previously observed in this compound. The observed scaling exponent $\gamma = 0.3 \pm 0.1$, independent of annealing. Fits of the stretched-exponential relaxation function~$G(t) = \exp[-(\Lambda t)^K]$ to the data yielded stretching exponentials~$K < 1$ for all samples. Annealed samples exhibited a reduction of the relaxation rate at low temperatures, indicating that annealing shifts fluctuation noise power to higher frequencies. There was no tendency of the inhomogeneous spread in rates to decrease with annealing, which modifies but does not eliminate the glassy spin dynamics reported previously in this compound. The correlation with residual resistivity previously observed for a number of NFL heavy-electron materials is also found in the present work.
\end{abstract}

\end{frontmatter}

\section{Introduction}
Muon spin relaxation ($\mu$SR) experiments in a number of $f$-electron non-Fermi liquid compounds and alloys have revealed anomalous behavior in the low-frequency spectrum of thermally-excited spin fluctuations in these materials~\cite{MHBI04}. The noise spectrum exhibits divergent behavior at low frequencies, with an increase in low-frequency noise power that correlates with increasing disorder as measured by the residual resistivity. Specifically, the longitudinal-field $\mu$SR (LF-$\mu$SR) relaxation function~$G(t,H)$ obeys a time-field scaling relation~$G(t,H) = G(t/H^\gamma)$, which is indicative of a power-law autocorrelation function~$\langle S(t)S(0)\rangle \propto t^{1-\gamma}$ for the $f$-ion spins \cite{KMCL96}. This behavior was previously observed in spin glasses above the freezing temperature~$T_g$~\cite{KMCL96,KGCB02} and in a number of NFL systems~\cite{MHBI04,MBHN01,MHBN02,MRYB03}. However, in the NFL systems studied to date there is no spin freezing down to low temperatures. 

One such disordered NFL compound is UCu$_4$Pd, which in its as-cast (unannealed) form exhibits a large residual resistivity (300--$400\ \mu\Omega$ cm) and strongly inhomogeneous static NMR and $\mu$SR Knight shifts~\cite{BMLA95} and relaxation rates \cite{MBHN01}. In UCu$_4$Pd $\mu$SR relaxation~\cite{MBHN01,MHBN02} and inelastic neutron scattering~\cite{AORL95b,AOCM01} values of the scaling exponent~$\gamma$ are found to agree ($\gamma \approx 0.3$), indicating a remarkable scaling of the spin dynamics scale over three orders of magnitude in frequency. EXAFS studies of UCu$_4$Pd~\cite{BMHC98} showed that Cu-Pd site exchange was responsible for much of the observed disorder. Subsequently, thermodynamic and especially transport properties of UCu$_4$Pd were found to depend strongly on annealing~\cite{WKKS00}, which EXAFS measurements~\cite{BSKW02} showed decreased the degree of site-exchange disorder.

\section{Experimental Results}
We have carried out LF-$\mu$SR experiments in unannealed and annealed samples of the heavy-fermion compound UCu$_4$Pd to study the effect of disorder on non-Fermi liquid behavior in this material. The measurements were carried out using the Low Temperature Facility at the Paul Scherrer Institute, Villigen, Switzerland. Data were taken on an unannealed sample and on samples annealed for 1 and 2 weeks at $750^{\circ}$C\@.

Time-field scaling was observed for all samples. An example is given in Fig.~1, which shows a scaling plot of $G(t,H)$ vs $t/H^\gamma$ at 50 mK for the 2-week annealed sample.
 \begin{figure}[ht]
     \begin{center}
     \includegraphics[width=0.45\textwidth]{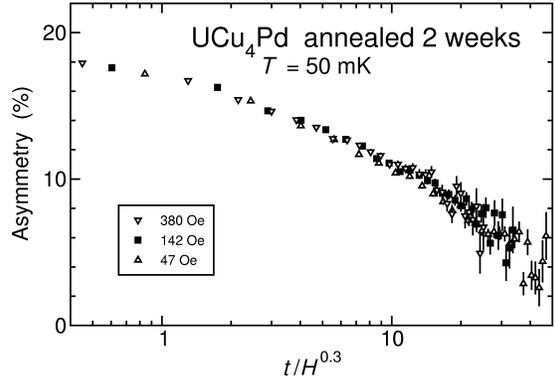}
     \end{center}
     \caption{Scaling plot of asymmetry relaxation function~$G(t,H)$ versus scaling variable~$t/H^\gamma$,$\gamma = 0.03$, in UCu$_4$Pd, unannealed and annealed for 1 and 2 weeks.}
 \end{figure}
The value of the scaling exponent for best overlap of data is $\gamma = 0.3 \pm 0.1$, close to the value 0.35 found previously from experiments on an unannealed sample~\cite{MBHN01}.

Fits to the LF-$\mu$SR data were made using the stretched-exponential relaxation function~$G(t) = \exp[-(\Lambda t)^K]$, where $\Lambda$ is a characteristic muon spin relaxation rate ($1/\Lambda$ is the time for decay of $G(t)$ to $1/e$ of its initial value), and $K < 1$ parameterizes the inhomogeneous spread in relaxation rates. Figure~2 shows the temperature and field dependence of $\Lambda$ from these fits.
 \begin{figure}[ht]
     \begin{center}
     \includegraphics[width=0.45\textwidth]{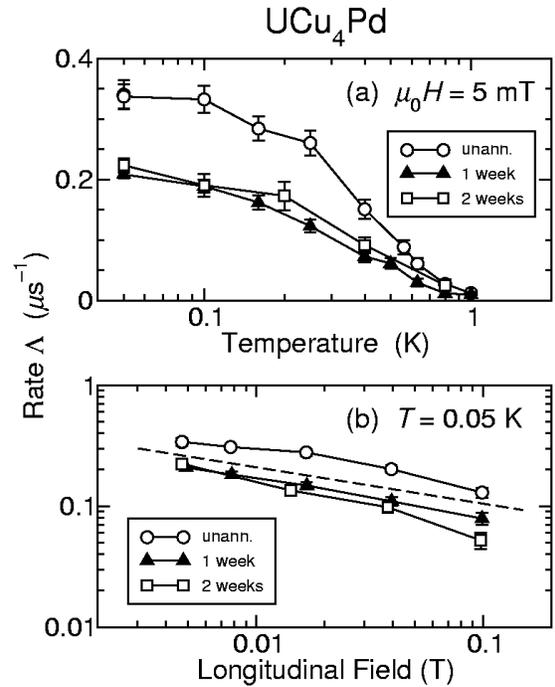}
     \end{center}
     \caption{Temperature and field dependence of relaxation rate~$\Lambda$ from fits of stretched-exponential relaxation function~$G(t) = \exp[-(\Lambda t)^K]$ in UCu$_4$Pd, unannealed and annealed for 1 and 2 weeks. The dashed line of slope $-0.3$ is shown for comparison.}
 \end{figure} 
Both the 1-week and 2-week annealed samples exhibit a reduction of $\Lambda$ at low temperatures, indicating that annealing shifts fluctuation noise power to higher frequencies. This behavior is expected from the previously-observed correlation between low-temperature relaxation and residual resistivity~\cite{MHBI04}. 

The field dependence, given in Fig.~2(b) for $T = 0.05$~K, shows that $\Lambda$ is roughly proportional to $H^{-\gamma}$, $\gamma = 0.3$, independent of annealing. Such a relation follows from time-field scaling if $G(t,H) = G[\Lambda(H)t]$, as is the case for the stretched exponential. The fact that $\gamma$ does not depend on annealing suggests that the degree of disorder does not significantly modify the low-frequency divergence of the fluctuation noise spectrum responsible for the muon spin relaxation.

The temperature and field dependence of the stretching exponential~$K$ are given in Fig.~3.
  \begin{figure}[ht]
     \begin{center}
     \includegraphics[width=0.45\textwidth]{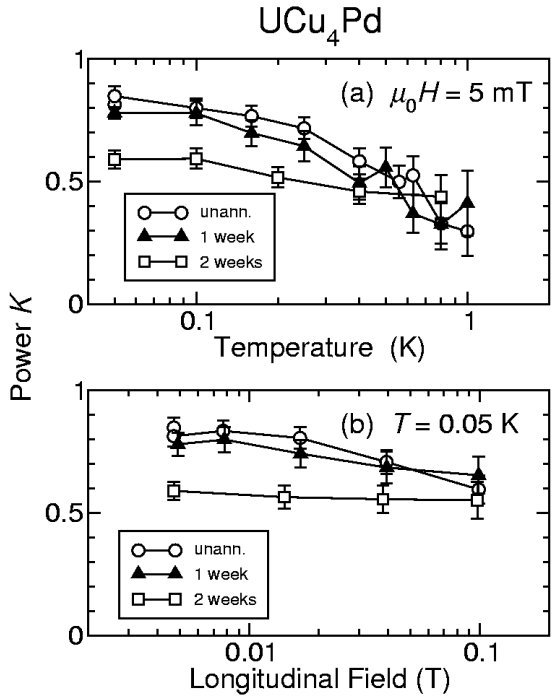}
     \end{center}
     \caption{Temperature and field dependence of stretching exponential~$K$ from fits of stretched-exponential relaxation function~$G(t) = \exp[-(\Lambda t)^K]$ in UCu$_4$Pd, unannealed and annealed for 1 and 2 weeks.}
 \end{figure}
Surprisingly, there is no tendency of $K$ to increase with annealing towards the value $K = 1$ expected in an ordered sample; indeed, Fig.~3 shows that $K$ decreases (and becomes more nearly independent of both temperature and field) in the 2-week annealed sample. This behavior indicates that while annealing shifts fluctuation noise power to higher frequencies it does so in an inhomogeneous manner that increases the spread in relaxation rates. As previously noted~\cite{MBHN01}, the tendency of $K$ to increase towards 1 at low temperatures may indicate more efficient averaging over the disorder, associated with an increasing correlation length as a low- or zero-temperature phase transition is approached.

\section{Conclusions}
Our LF-$\mu$SR results indicate that the glassy spin dynamics reported previously in this alloy\cite{MBHN01} are modified but not eliminated by annealing. (Here the term `glassy spin dynamics' designates slow paramagnetic-state fluctuations where the transfer of spectral weight to low frequencies is due to random and/or frustrated interactions between spins.)  The correlation with residual resistivity previously observed for a number of NFL heavy-electron materials is also found in the present work. As previously found from comparison of a number of different NFL systems~\cite{MHBI04,MRYB03}, low-frequency divergences characterize NFL spin fluctuations independently of the degree of disorder with, however, a strong increase of low-frequency fluctuation noise power in disordered NFL materials.

\section*{Acknowledgment}
We are grateful to A. Amato, C. Baines, and D. Herlach for their help during the experiments. One of us (D.E.M.) wishes to thank K. Ishida, Y. Maeno, and the Quantum Materials Group at Kyoto University, and also K. Nagamine and the MSL group at KEK, for their hospitality during visits when part of this work was carried out. These experiments were performed at the Swiss Muon Source, Paul Scherrer Institute, Villigen, Switzerland. The work was supported by US NSF Grants DMR-0102293 (Riverside), DMR-0203524 (Los Angeles),and DMR-0335173 (San Diego), by the Netherlands NWO and FOM (Leiden), and by the U.S. DOE, Contract FG02-04ER46105 (San Diego). Work at Los Alamos was carried out under the auspices of the U.S. DOE.



\end{document}